\documentstyle[preprint,prb,aps]{revtex}
\begin{document}
\draft
\title{Quantum phase transition in a multi-component \\
Bose-Einstein condensate in optical lattices}

\author{Guang-Hong Chen and Yong-Shi Wu}
\address{Department of Physics, University of Utah, 
Salt Lake City, Utah  84112}

\maketitle
\begin{abstract}
\baselineskip=0.90cm
We present the general lattice model for a multi-component
atomic Bose-Einstein system in an optical lattice. Using 
the model, we analytically study the quantum phase transition
between Mott insulator and superfluid. A mean-field theory
is developed from the Mott insulator ground state. When the
inter-species  interactions are strong enough, the Mott
insulator demonstrates the phase separation behavior. For
weak inter-species  interactions, the multi species system
is miscible. Finally, the phase diagram is discussed with 
the emphasis on the role of inter-species  interactions. 
The tips of the Mott insulator lobes do not depend on 
the inter-species  interactions, but the latter indeed 
modify the range of lobes. 
\end{abstract}


\newpage

\section{Introduction}
The study of quantum phase transitions (QPT) has attracted 
much interest in recent years \cite{rmp,sachdev}. The 
term ``quantum'' is used to emphasize that it is quantum 
fluctuations that play a vital role in driving the 
transition from one phase to another. In contrast, the 
usual thermodynamic phase transition at finite temperature 
is driven by thermal fluctuations which are experimentally 
controlled by tuning the temperature of the system. As 
temperature is lowered, the thermal fluctuations are 
suppressed and finally they are not strong enough to 
drive a finite temperature phase transition. However, 
this by no means implies that there would be no phase 
transition at very low temperature, since quantum 
fluctuations still exist and they may be sufficiently 
strong to drive a phase transition even at zero 
temperature. We call such a zero temperature phase 
transition a QPT, and it is experimentally accessible by
tuning parameters of the system other than temperature.

Several prominent examples have been extensively studied 
to demonstrate QPT. One example is quantum Hall (QH) 
systems, where different QH phases can be achieved 
by tuning either the magnetic field or carrier 
concentration\cite{rmp}. The second examle is a network
of Josephson junctions\cite{HD}. A Josephson junction is 
a tunnel junction connecting two superconducting metallic 
grains. A Cooper pair of electrons are able to tunnel back 
and forth between the grains. If the Cooper pairs can 
move freely from grain to grain in the network, the 
system is superconducting. However, since the grains 
are very small, it costs a charging energy to move a 
Cooper pair to neighboring grains. When the charging 
energy is big enough, the Cooper pairs fail to propagate 
among the grains and the network will be in an insulating 
phase. 

A third system that exhibits QPT involves the superfluid 
$ ^{4}He$. When the superfluid $ ^{4}He$ is absorbed 
in the porous media or on different substrates, the 
bosonic atoms in $ ^{4}He$ experience external forces 
from the other medium. When the interactions between 
atoms are much weaker than above external forces, the 
system is expected to be a superfluid. In the opposite 
limit, the superfluid phase can not be maintained, and 
the system will exhibit a Mott insulator behavior. 
Thus a superfluid-Mott-insulator phase transition is 
expected to happen, if one can tune the strength of 
atomic interactions. Detailed discussions can be found  
in Ref.\onlinecite{fisher} by Fisher et al. The 
starting point is the following boson Hubbard model:
\begin{equation}
\label{bH1}
H=-J\sum_{<i,j>}(a^{\dag}_ia_{j}+H.c.)
+\sum_{i}\varepsilon_in_i
+\frac{U}{2}\sum_{i}n_i(n_i-1).
\end{equation}  
Here $a_i$ and $a^{\dag}_{i}$ correspond to the 
bosonic annihilation and creation operators on the 
$i$-th lattice site, $n_i=a^{\dag}_ia_i$ the atomic 
number operator on the $i$-th site, and $\varepsilon_i$ 
the energy offset of the atom on the $i$-th site due 
to external harmonic confinement. The last term corresponds 
to the on-site repulsion between atoms, while the first 
term describes the tunneling of atoms between 
neighboring sites. At mean field level, starting with 
a strong coupling expansion, namely treating the hopping 
term as a perturbation, the system is found to have a 
QPT at the following critical 
value\cite{fisher,sheshadri,freericks,stoof1} for the 
ratio $U/J$:
\begin{equation}
\label{critical}
\frac{U}{J}=zn_0,
\end{equation}
where $z=2d$ for a $d$ dimensional simple lattice and 
$n_0$ is the inverse fraction of condensed atoms in a 
canonical ensemble. For instance, $n_0\approx 5.83$ 
for the three dimensional case. 

Experimentally, such critical point of QPT is very hard 
to access. Temperature is an annoying factor for a 
convincing demonstration of the QPT: The intrusion of 
thermal fluctuations often washes out the effects of 
quantum fluctuations. This makes the temperature window 
to observe the QPT small. Moreover, to make the system 
go cross the quantum critical point, we need to tune 
the controlling parameter carefully. In most of the 
studied cases, this is hard to manipulate. Even one 
can tune the parameter, the range of tunability is 
normally very small. Until very recently, in most 
cases only the magnetic field\cite{rmp,si} is the 
tunable parameter. Finally, the presence of disorder 
makes the observation of QPT even more difficult.  

Recently Ref.\onlinecite{markus} reported the success in 
realizing a superfluid-Mott-insulator phase transition 
in a gas of ultra-cold atoms in an optical lattice. 
This is a revolutionary breakthrough for experimental 
observation of a controllable QPT. They cooled the 
atomic gas of $ ^{87}Rb$ down to $10$ nK to 
realize the atomic Bose-Einstein condensation (BEC). 
Moreover, the BEC is loaded into a perfect, simple 
cubic, optical lattice formed by six criss-cross 
laser beams. By controlling the intensity of the 
laser beams, they can efficiently control the 
potential height of the above simple cubic lattice 
in a very large range. In addition, such a unique 
invention of the artificial lattice has the great 
advantage that the system is basically defect-free. 
By using this set-up, they successfully and repeatedly 
observed the QPT at the critical value 
given by Eq. (\ref{critical}). Thus an ideal playground 
for QPT  has been created in the atomic BEC system, which 
provides us an opportunity to test many theoretical 
predictions.

A significant difference between the atomic BEC 
superfluid and the $^{4}He$ superfluid is that the 
former allows atoms to condense with different 
internal states due to hyperfine splitting. This 
allows the order parameter of superfluid to possess 
a larger symmetry than the familiar $U(1)$\cite{ho1,ho2}. 
It is dubbed in the BEC community as spinor BEC. As 
pointed out by Ho\cite{ho1,ho2} and many others
\cite{stoof,zhou}, the spinor BEC possesses a whole 
host of quantum phenomena that are absent in the 
scalar cases: For instance, vector and quadrupolar 
spin wave modes, Skymions and other quantum orders 
etc. Experimentally, one can condense different 
matter species into one single internal state and 
study the effects of cross-species interactions. 
Throughout this paper, we would like to call such 
systems as multi-component BEC systems.

In the Mott insulator to superfluid quantum 
phase transition, quantum fluctuations and 
atomic interactions play a vital role. Without 
interactions, one has only the so-called band 
insulator. In the atomic gas, due to laser 
cooling technology, BEC can be realized 
simultaneously in several internal hyperfine 
levels\cite{chu}. This makes the experimental 
study of the  multi-component BEC possible. Among 
all  the interesting physics discovered in the 
multi-component BEC, the inter-species  
repulsive interactions play a very important 
role. Therefore, it would be very interesting 
to study how inter-species  interactions  affect 
the transition from the Mott  insulator  to 
superfluid.
 
Motivated both by experimental progress and 
by theoretical curiosity, we shall study in 
the present paper the superfluid-Mott-insulator 
transition in a multi-component BEC system in the 
presence of a periodic potential created by 
criss-cross laser beams. The layout of the paper
is the following: The boson-Hubbard model for the 
multi species is derived for the general case and
some special cases as well in Section II. In the 
section III, we study the ground state and its 
stability in the strong coupling limit. 
The phase boundary between superfluid and Mott 
insulator is determined for the two-component 
case in Section IV. Finally, we summarize our 
results in the Section V.

\section{The model}

\subsection{The general boson Hubbard model for a
multi-component BEC}
After including the optical lattice potential, 
the most general model Hamiltonian for a 
multi-component boson gas can be written, in 
the second-quantization notations, as
\begin{eqnarray}
\label{general}
H=\int d^3\vec{x}&\biggl[&\psi^{\dag}_i(x)
\biggl(-\frac{\hbar^2\nabla^2}{2m_a}
\delta_{ij}+U_{ij}(x)+V_i (x)\delta_{ij}
\biggr)\psi_j(x) \\ \nonumber
&+&\frac{g_{ij,kl}}{2}\psi^{\dag}_i(x)
\psi^{\dag}_j(x)\psi_k(x)\psi_l(x)\biggr],
\end{eqnarray}
where $m_a$ is the mass of an individual atom, the 
indices $i,j,k,l$ label the components of the atoms 
and the summation is assumed for repeated indices. 
Generically we allow the external potential $U_{ij}$ 
to have a non-diagonal part in the hyperfine spin 
basis, in which it represents a Josephson-type 
coupling between spin components\cite{wolfgang}. 
$V_i (x)$ denotes the optical lattice potential 
seen by atoms of species $i$. For the experimental 
configuration in Ref.\onlinecite{markus}, this 
lattice is modeled by
\begin{equation}
\label{lattice}
V(x,y,z)=V_0(\sin^2{kx}+\sin^2{ky}+\sin^2{kz}),
\end{equation} 
with $k$ the wave vector of the laser light 
and $V_0$ the depth of the potential well. In 
the multi-component case, the depth $V_{0,i}$
may depend on the species index $i$. The 
inter-atomic interactions in Eq. (\ref{general}) 
have been approximated as a contact interaction 
in which the coefficients $g_{ij,kl}$ describe 
the strength of various elastic and inelastic 
collisions.

For a single atom in the trap and the periodic 
potential, the energy eigenstates are Bloch states. 
In the tight-binding (TBA) limit, we can superpose 
the Bloch states to get a set of Wannier functions,
which are localized on individual lattice 
sites. Within the single band approximation, we can 
expand the field operators in the Wannier basis as
\begin{equation}
\label{wannier1}
\psi_i(x)=\sum_{n}b_{ni}w_{i}(x-x_n),
\end{equation}
where $w_{i}(x-x_n)$ is the Wannier function around 
lattice site $n$. Using Eq. (\ref{wannier1}), the 
general Hamiltonian (\ref{general}) is reduced to 
a generalized boson Hubbard Hamiltonian for the 
multi-component BEC:
\begin{equation}
\label{bH2}
H=-\sum_{<m,n>}J^{ij}_{mn}(b^{\dag}_{mi}b_{nj}+H.c.)
+\sum_{m}\varepsilon_{mi}b^{\dag}_{mi}b_{mi}
+\frac{U_{ij,kl}}{2}\sum_mb^{\dag}_{mi}
b^{\dag}_{mj}b_{mk}b_{ml}.
\end{equation}
Here $J^{ij}_{mn}$ is the hopping matrix element 
between two adjacent lattice sites $m$ and $n$. 
It is defined by
\begin{equation}
\label{hopping}
J^{ij}_{mn}=-\int d^3x w^{*}_{i}(x-x_m)
[-\frac{\hbar^2\nabla^2}{2m_a}
\delta_{ij}+V_i \delta_{ij}+U_{ij}-
\frac{1}{2}(U_{ii}+U_{jj})]w_j(x-x_n).
\end{equation}
Within the Hubbard approximation, the hopping 
integral is lattice site independent, i.e., 
$J^{ij}_{mn}\approx J^{ij}$. $\varepsilon_{ni}$ 
describes the energy offset on each site due to
the trap confinement. It is defined as
\begin{equation}
\label{offset}
\varepsilon_{ni}=\int d^3xU_{ii}(x)|w_{i}(x-x_n)|^2.
\end{equation}
(Here we only consider the lowest band in the optical 
lattice, whose bottom is taken to be the zero point 
for energy.) Finally, to get the on-site interactions 
in Eq. (\ref{bH2}), the Hubbard approximation has been 
used to approximate the multi-center integral 
as a single-center one; namely, we have
\begin{eqnarray} 
\label{hubbard}
U_{ij,kl}&=&g_{ij,kl}\int d^3x w_i^{*}(x-x_{n1})
w_j^{*}(x-x_{n2})w_k(x-x_{n3})w_l(x-x_{n4})
\\ \nonumber
&\approx&g_{ij,kl}\int d^3x w_i^{*}(x)w_j^{*}(x)
w_k(x)w_l(x).
\end{eqnarray}

The general form of the boson Hubbard model (\ref{bH2}) 
contains a large number of parameters. In the following, 
we would like to discuss several special cases which 
might be relevant to experiments. The first simple case
is, of course, given by Eq. (\ref{bH1}) for a single 
component BEC.  It is first derived and studied in the 
context of  atomic BEC in Ref. \onlinecite{zoller}. 

\subsection{Two-component boson Hubbard model}

The second example we will discuss is the two-component 
BEC. Experimentally, the simultaneous condensation of 
$^{87}Rb$ atoms in the two internal states ($F=2, M=2$) 
and ($F=2, M=-1$) has been accomplished by Myatt et 
al.\cite{myatt}. For this case, we discuss Bose condensed 
atoms with two internal hyperfine levels $|A>$ and $|B>$. 
The atoms interact only through the following three 
channels: $AA$, $BB$, and $AB$ type elastic collisions. 
Then our Hamiltonian (\ref{bH2}) is reduced to
\begin{eqnarray}
\label{bH3}
H_2=&-&\sum_{<m,n>}\biggl[J^{A}b^{\dag}_{mA}b_{nA}
+J^{B}b^{\dag}_{mB}b_{nB}
+J^{AB}b^{\dag}_{mA}b_{nB}+H.c.\biggr] \\ \nonumber
&+&\sum_{m}\biggl(\varepsilon_{mA}n_{mA}
+\varepsilon_{mB}n_{mB}\biggr) 
\\ \nonumber
&+&\frac{1}{2}\sum_m\biggl[U_An_{mA}(n_{mA}-1)
+U_Bn_{mB}(n_{mB}-1) +U_{AB}n_{mA}n_{mB}\biggr].
\end{eqnarray}
A similar energy-level and interaction pattern 
has been discussed by another group in a different 
context\cite{cirac}. To be concrete, we 
focus on the situation in which the trap potential 
is diagonal in internal space, namely,  
$U_{ij}$ only have diagonal components $U_{ii}$. In 
this case, it follows from Eq. (\ref{hopping}) that
\begin{equation}
\label{jab}
J^{AB}_{ij}=0.
\end{equation}
To get more insight into the parameters in Eq.
(\ref{bH3}), we have to use the explicit form of 
the Wannier functions. To do so, we notice that 
the optical lattice potential is sinusoidal. 
The Wannier function could be constructed as the 
localized one determined by the following 
eigenvalue problem:
\begin{equation}
\label{eigen}
[\frac{P^2}{2m_a}+V(x,y,z)]\phi(x,y,z)
=E\phi(x,y,z).
\end{equation}
The lattice sites are given by minima of the 
lattice potential $V(x,y,z)$; around them 
the potential $V$ is approximately quadratic:
\begin{equation}
\label{approc}
V(x,y,z)\approx\frac{1}{2}m_a
\omega(x^2+y^2+z^2), 
\end{equation}
where $\omega$ is given by
\begin{equation}
\label{omega}
\omega=2m_ak^2V_0.
\end{equation}
(Here for simplicity, we assume that the depth of
the optical potential is the same for different 
species; it is straightforward to generalize our 
results to the case with $V_0\to V_{0,i}$ dependent 
on the species index $i$.) 
Therefore, within the single band approximation, 
the Wannier function is approximately given by 
the ground state wavefunction of a three-dimensional 
harmonic oscillator. Namely
\begin{equation}
\label{wannier}
w_{A/B}=\biggl(\sqrt{\frac{\alpha}{\sqrt{\pi}}}
\biggr)^3 \exp[-\frac{1}{2}\alpha^2(x^2+y^2+z^2)],
\end{equation}
where $\alpha=\sqrt{\frac{m_a\omega}{\hbar}}$. 
Noting that the Wannier functions are independent 
of species indices within our approximations. 
Thus we can take
\begin{equation}
\label{para}
J_A=J_B=J.
\end{equation}
Finally, the on-site energy is the original 
inter-atomic interaction with an extra numerical 
factor $\int [w(x,y,z)]^4$. Therefore, the 
Hamiltonian (\ref{bH3}) can be cast into  
\begin{eqnarray}
\label{bH3'}
H_2=&-&J\sum_{<m,n>}\biggl(b^{\dag}_{mA}b_{nA}
+b^{\dag}_{mB}b_{nB}+H.c.\biggr)
+\sum_{m}\biggl(\varepsilon_{mA}n_{mA}
+\varepsilon_{mB}n_{mB}\biggr) 
\\ \nonumber
&+&\frac{1}{2}\sum_m\biggl[U_An_{mA}(n_{mA}-1)
+U_Bn_{mB}(n_{mB}-1)+U_{AB}n_{mA}n_{mB}\biggr].
\end{eqnarray}

Our Hamiltonian (\ref{bH3'}) is different from the 
two-species boson Hubbard model proposed in Ref. 
\onlinecite{zoller}, where the authors assumed that 
two species $A$ and $B$ are placed in two different 
optical lattices with a relative half-period 
shift. Also a drive laser has been applied 
to induce the transition between species $A$ and $B$. 
In this situation, $J^{A}$ and $J^{B}$ in (\ref{bH3}) 
should be neglected since they represent the next 
nearest neighbor hopping. Moreover, the on-site 
mutual interactions between two species are of higher 
orders compared with the on-site interactions for 
the same species. In this way, we recover their 
Hamiltonian\cite{zoller}
\begin{eqnarray}
\label{bH3''}
\tilde{H}_2=&-&J\sum_{<m,n>}\biggl(b^{\dag}_{mA}b_{nB}
+H.c.\biggr) +\sum_{m}\biggl(\varepsilon_{mA}n_{mA}
+\varepsilon_{mB}n_{mB}\biggr) \\ \nonumber
&+&\frac{1}{2}\sum_m\biggl[U_An_{mA}(n_{mA}-1)
+U_Bn_{mB}(n_{mB}-1)+U_{AB}n_{mA}n_{mB}\biggr].
\end{eqnarray}

\subsection{Boson Hubbard model for spinor BEC}

Another well studied example of the multi-component BEC 
is the so-called spinor BEC\cite{ho2,wolfgang}. For 
a system of spin $f=1$ bosons, such as $^{23}Na$, 
$^{39}K$, and $^{87}Rb$ atoms, the form of the 
inter-atomic interactions is largely constrained 
by symmetries. In this case, the number of the 
interaction parameters are reduced to two. The 
interaction potential can be written as
\begin{equation}
\label{spinor}
V_{int}(x_1,x_2)=(g_0+g_2\vec{F}_1\cdot\vec{F}_2)
\delta(x_1-x_2),
\end{equation}
where the parameters $g_0$ and $g_2$ are defined 
by the scattering length $a_2$ and $a_0$ as
\begin{eqnarray}
\label{g1}
g_0&=&\frac{4\pi\hbar^2}{m_a}\frac{2a_2+a_0}{3}; \\
\label{g2} 
g_2&=&\frac{4\pi\hbar^2}{m_a}\frac{a_2-a_0}{3}.
\end{eqnarray}
To facilitate the discussion, we choose a 
basis to make the trapping potential $U_{ij}$ 
in Eq. (\ref{general}) diagonal. Thus the hopping
integral is non-vanishing only between the
same species. In addition, the on-site interactions 
are reduced to the following eight terms:
\begin{equation}
\label{spinorinteraction}
H_{int}=\frac{1}{2}\sum_{mA}U_{A}n_{mA}(n_{mA}-1)
+\frac{1}{2}\sum_{m,A\neq B}U_{AB}n_{mA}n_{mB}
+\frac{U_0}{2}\sum_{m}(b^{\dag 2}_{m0}
b_{m1}b_{m\bar{1}}+H.c).
\end{equation}
Here the species index $A=1,0,\bar{1}(=-1)$. $U_A$, 
$U_{AB}$, and $U_0$ are determined by the parameters 
$g_0$, $g_2$ and the Wannier functions. In particular, 
the parameter $U_0$ is proportional to $g_2$. 
Moreover, $g_2$ is determined by the difference 
between scattering lengths as shown in Eq. (\ref{g2}). 
For the sodium case, the difference between two 
scattering lengths is very small ($0.29nm$) compared 
with $2a_2+a_0=7.96nm$. Therefore, $g_2 \ll g_0$ so 
that we can neglect the spin relaxation channel in 
the interaction terms. Namely, we set $U_0=0$ as 
the zeroth order approximation. Within this 
approximation, we get the following boson Hubbard 
model for a spin-$1$ spinor BEC, when it is loaded 
into the optical lattice potential:
\begin{eqnarray}
\label{spinorBEC}
H_3=&-&\sum_{<m,n>A}(J^{A}b^{\dag}_{mA}b_{nA}+H.c.)
+\sum_{mA}\varepsilon_{mA}n_{mA}  \\ \nonumber
&+&\frac{1}{2}\sum_{mA}U_{A}n_{mA}(n_{mA}-1)
+\frac{1}{2}\sum_{m,A\neq B}U_{AB}n_{mA}n_{mB}.
\end{eqnarray}
In the next section, we will discuss the possible 
mean field phase diagram for the superfluid-Mott 
insulator phase transition by starting with the
two-component boson Hubbard Hamiltonian (\ref{bH3}).


\section{Mott ground state and its stability}

We are going to employ the strong coupling expansion 
to develop a mean field theory. In the strong 
coupling limit, the hopping term can be treated as 
a perturbation. In the zeroth order approximation, 
we ignore it for a while. The Hamiltonian is then 
decoupled  for the site index. The ground state is 
given by the  occupation number state $|n_A,n_B>$, 
with the  wavefunction 
\begin{equation}
\label{ground}
|Gnd>_{MF}\sim \prod_{m}(b^{\dag}_{mA})^{n_A}
(b^{\dag}_{mB})^{n_B}|0>.
\end{equation}
To get the ground state energy, we need to minimize 
the energy at each site (for this purpose, we neglect 
the site index in the following discussions). Namely, 
we need to minimize the energy function $E(n_A,n_B)$ 
given by
\begin{equation}
\label{energy}
E(n_A,n_B)=\varepsilon_A n_A+\varepsilon_B n_B+
\frac{1}{2}[U_An_A(n_A-1)
+U_Bn_B(n_B-1)+U_{AB}n_An_B].
\end{equation}
If we skip over the fact for the moment that the 
occupation number $n_A$ and $n_B$ are integers, 
then the condition to minimize the above energy 
function are given by
\begin{eqnarray}
\label{condition1}
U_An_A+U_{AB}n_B&=&\frac{U_{A}}{2}-\varepsilon_A, \\ 
U_{AB}n_A+U_{B}n_B&=&\frac{U_{B}}{2}-\varepsilon_B.
\end{eqnarray}
Sovling the two coupled linear equations, we get
\begin{eqnarray}
\label{solution1}
n_A&=&\frac{U_B(U_A-U_{AB})+2(\varepsilon_BU_{AB}
-\varepsilon_AU_B)}{2(U_AU_B-U^2_{AB})}, \\ 
n_B&=&\frac{U_A(U_B-U_{AB})+2(\varepsilon_AU_{AB}
-\varepsilon_BU_A)}{2(U_AU_B-U^2_{AB})}.
\end{eqnarray}
Now we take care of the fact that the occupation 
numbers must be integer. So the actual numbers 
to minimize the energy are the two integers closest 
to the above $n_{A/B}$. To do so, we can write 
$n_{A/B}$ in terms of the closest integer numbers 
$n^{0}_{A/B}$ and the decimal parts, i.e.
\begin{equation}
\label{decimal}
n_A=n^{0}_A+\alpha,\hspace{1.0cm} 
n_B=n^{0}_B+\beta,
\end{equation} 
where the numbers $\alpha$ and $\beta$ 
satisfy
\begin{eqnarray}
\label{alphabeta}
-\frac{1}{2}<\alpha=n_A-n^{0}_A<\frac{1}{2}, \\ 
-\frac{1}{2}<\beta=n_B-n^{0}_B<\frac{1}{2}.
\end{eqnarray}
Namely, when the parameters of the system 
satisfy the following conditions
\begin{eqnarray}
\label{condition2}
n^{0}_A-1&<&\frac{U_{AB}(U_{AB}-U_{B}+2\varepsilon_B)
-2\varepsilon_AU_B}{2(U_AU_B-U^2_{AB})}<n^{0}_A, \\ 
n^{0}_B-1&<&\frac{U_{AB}(U_{AB}-U_{A}+2\varepsilon_A)
-2\varepsilon_BU_A}{2(U_AU_B-U^2_{AB})}<n^{0}_B,
\end{eqnarray}
the occupation numbers $(n^{0}_A, n^{0}_B)$ 
minimize the energy $E(n_A,n_B)$. 

One loose end in above discussions is that we have 
assumed the minimal occupation numbers are non-zero. 
If one of the occupation numbers is zero, then it 
means that our ground state is not stable due to 
the mutual interactions between different species. 
To get the stability condition for the uniform 
ground state, we need to diagonalize the interaction 
terms, namely the following quadratic form:
\begin{equation}
\label{interaction}
U(n_A,n_B)=\frac{1}{2}(U_An^2_A
+U_Bn^2_B+U_{AB}n_An_B).
\end{equation}
The eigenvalues of this quadratic form are 
\begin{eqnarray}
\label{eigen2}
n_{\pm}&=&\frac{1}{4}[(U_A+U_B)\pm
\sqrt{(U_A-U_B)^2+U^2_{AB}}] \\ \nonumber
&=&\frac{1}{4}[(U_A+U_B)\pm
\sqrt{(U_A+U_B)^2+(U^2_{AB}-4U_AU_B)}].
\end{eqnarray}
Therefore, $n_{-}$ may become negative; if 
so, the interaction manifold is saddle-like 
and one cannot really minimize the ground 
state energy with two non-zero occupation 
numbers. Thus, in one spatial region, one 
of the species must have zero occupation. 
In other words, the ground state of the 
system must be phase separated, when the 
following condition is satisfied:
\begin{equation}
\label{phaseseperation}
U_AU_B\ge \frac{1}{4}U_{AB}.
\end{equation}
This condition (\ref{phaseseperation}) for 
phase separation is analogous to that of an
ordinary two-component BEC (without being 
loaded into an optical lattice)\cite{chui}. 
In the case when the Wannier functions are
the same for both species, this condition 
is reduced precisely to the one in the 
absence of the optical lattice.

\section{Phase transition to superfluid}

In this section, we are going to present a 
mean-field theory based on the ground state 
developed in the preceding section. The 
hopping processes correspond to moving bosons 
from one site to another. This process allows
bosons at different sites communicate with 
each other and finally they conspire to 
establish macroscopic coherence under 
appropriate conditions. In this way the system 
can enter a superfluid state with indefinite 
filling of bosons at each site.

The consistent mean-field theory we shall use 
corresponds to the following decomposition of 
the hopping terms:
\begin{eqnarray}
\label{decomposition}
b^{\dag}_{m}b_{n}&\approx&<b^{\dag}_m>b_{n}
+b^{\dag}_m<b_n>-<b^{\dag}_m><b_n>, 
\\ \nonumber
&=&\phi(b^{\dag}_m+b_{n})-\phi^2,
\end{eqnarray} 
where $\phi=<b^{\dag}_m>=<b_n>$ is the superfluid 
order parameter. In the case at hand, we have 
taken the order parameter to be real. In this 
decomposition, the higher order fluctuations
$(b^{\dag}_m-\phi)(b_n-\phi)$ have been 
neglected. It reflects the fact that in the 
ground state energy corrections we neglect the
correlation energy. 
Generally speaking, this process will increase 
the energy of the system; however, when the system 
parameters satisfy certain conditions, this process 
will not cost any energy or even will lower the 
energy of the system. This signals the occurrence
of a phase transition. Therefore, the vanishing
energy correction due to the hopping process should
give us the phase boundary. In the following, we 
shall determine the phase boundary using second 
order perturbation theory.

The resulting mean-field version of the hopping 
Hamiltonian can be written as
\begin{eqnarray}
\label{mean}
H^{eff}&=&\sum_{m}H^{eff}_m  \\ \nonumber
&=&-zJ\sum_{m}\biggl[\phi_A(b^{\dag}_{mA}+b_{mA}) 
+\phi_B(b^{\dag}_{mB}+b_{mb})
-(\phi^2_A+\phi^2_B)\biggr].
\end{eqnarray}
Here $z$ is the number of nearest-neighbor sites. 
Since it is a single sum over all lattice sites, 
we drop the site index from now on.

The first order correction to the energy vanishes, 
due to the fact that the ground state is a product
of number eigenstates at each site, and thus the 
average of an annihilation or creation operator is 
just zero. The second order correction to the 
energy is given by the following well-known 
expression:
\begin{equation}
\label{second2}
E^{(2)}_g=\sum_{n\neq g}\frac{|<g|H^{eff}_m|n>|^2}
{E^{(0)}_g-E^{(0)}_{n}},
\end{equation}
where $|n>=|n_A, n_B>$ denotes the unperturbed 
state with $n_A$ and $n_B$ atoms for each species,
respectively. Correspondingly, $|g>=|n^{0}_A, 
n^{0}_B>$ is the ground state and the occupation 
numbers are given by Eq. (\ref{condition2}). 
A straightforward calculation gives the second 
order correction to the ground state energy 
as follows:
\begin{eqnarray}
\label{second}
E^{(2)}_g&=&J^2z^2\phi^2_A\biggr[\frac{n^{0}_A}
{\varepsilon_A+U_A(n^{0}_A-1)+U_{AB}n^{0}_B/2}+
\frac{n^{0}_A+1}{-\varepsilon_A-U_An^{0}_A
-U_{AB}n^{0}_B/2}\biggl] 
\\ \nonumber
&+&J^2z^2\phi^2_B\biggr[\frac{n^{0}_B}
{\varepsilon_B+U_B(n^{0}_B-1)+U_{AB}n^{0}_A/2}
+\frac{n^{0}_B+1}{-\varepsilon_B-U_Bn^{0}_B
-U_{AB}n^{0}_A/2}\biggl] \\ \nonumber
&+&Jz(\phi^2_A+\phi^2_B).
\end{eqnarray}
Therefore, the phase boundaries between the 
Mott insulator and the superfluid for 
species $A$ and $B$, respectively, are given 
by the following conditions:
\begin{eqnarray}
\label{supercondition}
1+Jz\biggr[\frac{n^{0}_A}{\varepsilon_A
+U_A(n^{0}_A-1)+U_{AB}n^{0}_B/2}+
\frac{n^{0}_A+1}{-\varepsilon_A-U_An^{0}_A
-U_{AB}n^{0}_B/2}\biggl]
&=&0,  \\ \nonumber
1+Jz\biggr[\frac{n^{0}_B}{\varepsilon_B
+U_B(n^{0}_B-1)+U_{AB}n^{0}_A/2}
+\frac{n^{0}_B+1}{-\varepsilon_B
-U_Bn^{0}_B-U_{AB}n^{0}_A/2}\biggl]
&=&0.
\end{eqnarray}
Solving above equations yields
\begin{eqnarray}
\label{chemical}
\varepsilon^{\pm}_A=-\frac{1}{2}\biggl[
U_{AB}n^0_B+U_A(2n^0_A-1)-Jz
\pm\sqrt{U^2_A-2U_AJz(2n^0_A+1)
+(Jz)^2}\biggr], \\ \nonumber
\varepsilon^{\pm}_B=-\frac{1}{2}\biggl[
U_{AB}n^0_A+U_B(2n^0_B-1)-Jz
\pm\sqrt{U^2_B-2U_BJz(2n^0_B+1)
+(Jz)^2}\biggr].
\end{eqnarray}
When the chemical potential $\varepsilon_A$ 
is in the following region 
\begin{equation}
\label{che1}
\varepsilon^{-}_A<\varepsilon_A
<\varepsilon^{+}_A,
\end{equation}
the energy correction due to the tunneling 
events of species $A$ is positive and thus 
the Mott-insulator is the stable ground state. 
A similar result is valid for species $B$. 
When the inter-species interactions $U_{AB}$ 
and the intra-species interactions $U_{A/B}$ 
satisfy the condition (\ref{phaseseperation}), 
we see that the mutual interactions and 
non-zero occupation for the other species 
indeed modify the phase boundary for both 
species. When the mutual-species repulsion
is strong enough to make the system phase 
separated, then the system can be viewed as 
two totally independent single species in 
the optical lattice. However, the most 
interesting observation from Eq. (\ref{chemical}) 
should be that the tips of the lobes given by 
condition (\ref{che1}) is independent of the
inter-species interaction $U_{AB}$ and thus 
it is independent of the occupation number 
of the other species too! The tips of the 
lobes for species $A$ and $B$ are given by
\begin{equation}
\label{tips}
\biggl(\frac{U_{A/B}}{Jz}\biggr)_c
=2n^{0}_{A/B}+1+\sqrt{(2n^{0}_{A/B}+1)^2-1},
\end{equation}
which is the same as for the single species 
boson Hubbard model. The tip of the first 
lobe is given by the critical value 
$U_{A/B}/(Jz)=5.83$ for both species. It is 
worth noting that the above result is valid 
even if the hopping integrals are not equal: 
$J_A\neq J_B$; the only thing we need to do 
is to scale $U_{A/B}$ by the corresponding 
hopping parameter $J_{A/B}$ respectively.

\section{conclusions and discussions}

In this paper, we have analytically studied the 
quantum phase transition between superfluid and
Mott insulator for a multi-component BEC system
in an optical lattice. 
Theoretically, this is a generalization of the 
well-studied case of the boson Hubbard model in 
the condensed matter literature. Experimentally, 
the first beautiful observation of such a 
quantum phase transition is accomplished by 
loading a one-component atomic Bose-Einstein 
condensate into an artificial optical lattice and
thus it would be nice to study the role of
the inter-species interactions in the QPT for
the multi-component cases.

In the first part of the paper, we have 
generalized the single species boson Hubbard 
model to the multi-component case with most 
general interactions. To be concrete, we have 
reduced our general boson Hubbard model to 
the two- and three-component cases under 
appropriate conditions.

Starting with the two-component boson Hubbard 
model, we developed a mean-field theory to study 
the quantum phase transition. Depending on  
inter-species interactions, the system may be
in different ground states. If the repulsion 
between two species is not very strong, the two 
species can co-exist; namely, the system is 
miscible. However, if the this repulsion is 
sufficiently strong, the two species may become
immiscible, and the ground state will demonstrate 
the behavior of phase separation. Namely two 
species of Mott insulator will stay in separate
spatial regions. After turning on the tunneling 
terms, the ground state energy will get corrections 
from tunneling. We calculated the energy 
corrections up to second order and determined the 
boundary between the gain and loss in energy. We 
found that the inter-species interactions indeed 
can change the range of the parameters for the 
Mott insulator. However, the inter-species 
interactions can not 
change the position of the tips for the Mott 
insulator lobes. The phase diagram of the 
two-component boson Hubbard model also 
demonstrates a richer structure. From our
analytical treatment, we conclude that the 
following three different phases are possible: 
(1) Both species $A$ and $B$ are in the superfluid 
phases; (2) one of the spieces is still in the
superfluid phase, while the other is in the Mott 
insulator phase; (3) both species are in the Mott 
insulator phase. 

Finally, several further remarks are in order. 
(1) Above results for a two-component system 
can be directly generalized to the spinor BEC 
and other systems with more components. (2) One 
can check without difficulty that there is 
indeed an energy gap in the excitation spectrum
of the Mott insulator. The gap at zero momentum 
is determined by the on-site Coulomb energy $U_{A/B}$ 
and is independent of the inter-species interaction 
$U_{A/B}$. Right at the phase transition, due to 
the gain of tunneling, the energy gap closes. 
Therefore, the system becomes compressible 
(or gapless) and therefore in a superfluid phase.
(3) In this paper, our studies of the  phase diagram for 
the multi species boson-Hubbard model have been
restricted in the case where the Josephson-type
tunneling term can be neglected. Under certain
experimental conditions, such terms would be
dominant and the physics is significantly changed.
The results for this case will be published 
somewhere else\cite{chen-wu}.

Acknowledgments

This research was supported in part by the US 
National Science Foundation under Grants No. 
PHY-9970701.


\begin{figure}
\caption{The schematic phase diagram of the two-component boson-Hubbard model.
The phase diagram is divided into three regions for fixed occupation
numbers $n_{A/B}$. One region is superfluid phase for both components 
(S.F. phase); the second region is a mixture of superfluid phase for one
component with Mott insulator phase for the other (S.M. phase);
The third region is Mott insulator for both components (M.I. phase).}
\end {figure}

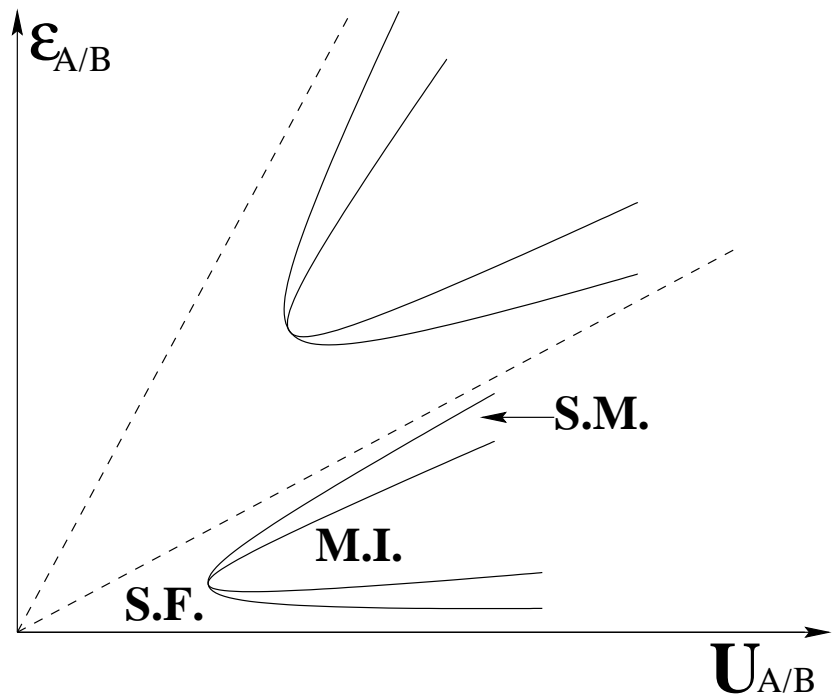
\end{document}